\documentstyle[epsf,a4,12pt]{article}
\title{Analytical modeling of large-angle CMBR anisotropies
from textures}
\author{\\ \\ J.C.R. Magueijo \\ \\
Mullard Radio Astronomy Observatory,
Cavendish Laboratory\\ Madingley Road\\ Cambridge, CB3 0HE, UK
\\ and \\  Department of Applied Mathematics and Theoretical
Physics\\University of Cambridge\\Cambridge CB3 9EW, UK.}
\date{}

\begin{document}
%XXX hack:
\renewcommand{\epsffile}[1]{Missing author supplied figures}
\bibliographystyle{unsrt}

\maketitle
\begin{abstract}
We propose an analytic method for predicting the
large angle CMBR temperature
fluctuations induced by model textures.
The model makes use of only a small number
of phenomenological parameters which ought to
be measured from  simple simulations.
We derive semi-analytically the $C^l$-spectrum
for $2\leq l\leq 30$ together with its associated
non-Gaussian cosmic
variance error bars. A slightly tilted spectrum with an extra
suppression at low $l$ is found, and we  investigate the dependence
of the tilt on the parameters of the model.
We also  produce a prediction for the two
point correlation function. We find a high level of cosmic confusion
between texture scenarios
and  standard inflationary theories in any of these quantities. However,
we discover that a distinctive non-Gaussian signal ought to be expected at
low $l$, reflecting the prominent effect of the last texture
in these multipoles.
\end{abstract}
\vspace{1cm}
\begin{flushright}
MRAO/1782 \\ DAMTP/94-101
\end{flushright}
\thispagestyle{empty}
\pagebreak

\pagestyle{plain}
\setcounter {page}{1}
\section{Introduction}
In the recent past, a large amount of work has been directed
towards predicting the CMBR temperature anisotropies associated with
inflationary scenarios (see \cite{lidl} for a review), and with the various
topological defect scenarios (e.g. \cite{txpen,shell}).
Inflationary scenarios are far better explored in this respect.
The main reason for this is that it is easier in these
models to plead ignorance of the detailed mechanism responsible
for the density fluctuations of the Universe. Quantum fluctuations
in the metric are a fuzzy subject, and hand-waving arguments
for Gaussian fluctuations with a particular type of spectrum
are normally considered acceptable. Once this is done, one is
free to take advantage of the well developed and sophisticated
industry of methods aimed at treating Gaussian fluctuations
\cite{efst}. A precise prediction
for the outcome of concrete experiments follows
easily, making inflation popular with experimentalists.
The state of affairs  in topological
defect scenarios is rather different. No lack of fundamental
physics understanding hampers working out in rigour the defect
network evolution and the fluctuations they induce on
the matter and radiation of the Universe. The price to pay
for our honesty is that a serious treatment of the problem
is computationally overpowering. A reliable link between
the defect network and the perturbations it induces is
undoubtedly still missing. To top off the trouble, it turns out
that the fluctuations induced by defects are often non-Gaussian.
Setting up a data-analysis framework geared towards non-Gaussian
fluctuations is virgin ground (but see
\cite{efst,coles,scar,falk} for Gaussianity tests).
Overall, even though defect scenarios
have so far been poor in solid predictions, they have a
potentially polemic value not to be discarded.
Non-Gaussian data analysis does
not exist partly because non-Gaussian theories are
too vast a class to be treated in one go. Defect scenarios
provide a motivated and topical example
with which one can rehearse non-Gaussian data analysis
techniques and experimental strategies.

The main technical complication in predicting CMBR fluctuations in
defect scenarios arises from the fact that ${\delta T\over T}$
maps photograph the whole sky (that is, our past
light cone clipped by the last scattering surface) and not just
its intersection with the last scattering surface. Hence the defect
network, surrounding matter, and CMBR photons have to be evolved
through a large number of expansion times. The problem is usually tackled
by means of full-blown simulations \cite{txpen,shell}. These
require access to supercomputers, and even so face severe
limitations. Important issues like cosmic variance in defect scenarios
have hardly been addressed. A possible computational short cut
can be found if one is
to assume scaling. The defects network is expected to scale,
by which one means that it looks statistically the same at any time,
once the network length scale is equated to the horizon distance.
This suggests dividing the sky cone into cells corresponding to
expansion times and horizon volumes. Scaling makes it
plausible that the CMBR pattern in each of these cells
is statistically the same, once some angular scaling procedure
is applied to it. This being the case, all that is required from the
simulations is good statistics for the CMBR spots induced
by defects inside horizon size boxes during one expansion
time. Such simulations are considerably easier to perform.
A prototype of this type
of study is \cite{borr}. In the case of textures, the short-cut
which we have described has been adopted by \cite{durr}
where a ``scaling-spot-throwing'' process was implemented
on a computer. The spots were all derived from model SSSS
(self-similar and spherically symmetric) collapses.
In this paper we advocate the use of analytical techniques
for scaling-spot-throwing. An analytical approach allows
the derivation of exact formulae for the $C^l$, $C(\theta)$,
and their cosmic variance, from which interesting general properties
can be derived. These properties are naturally less
evident in computer simulations, unless one explicitly sets out to
find them. Furthermore, an analytical approach allows the treatment of
a more general case than \cite{durr}. The method proposed here
applies to  quasi-circular spots with any profile.
The scaling size ($p_s$), number density ($n$), and brightness factor
($a$)
of the spots are also left as free parameters.
We should admit that only
the brightness factor is a truly free parameter in the
texture model. All others should be considered as
phenomenological in origin, eventually to be fixed
by simulations. As they still constitute a controversial
matter, we choose to consider a broader class of
model spots. We can then sort out which spot properties
it is really important to measure from the simulations.

The plan of this paper is as follows.
In Sec.\ref{ITT} we start by deriving the statistical properties
of scaling
texture model collapses as they appear in the sky cone, and from these
a formal solution for the joint distribution function of the CMBR
$a^l_m$ is given.
Using this solution, in Secs.\ref{apscl} and \ref{apsscl} we derive
exact formulae for the $C^l$ and $\sigma (C_l)$ in texture models.
We prove the general result that the cosmic variance in the  $C_l$
is larger in texture scenarios than in Gaussian theories.
The excess variance is negligible for high $l$, but significant at low
$l$. In Secs.\ref{apsprof} and \ref{apsclres} we compute
numerically the $C^l$ spectra (with cosmic variance error bars)
for a large class of spot types. We find that the spots' profile,
intensity, and number density affect only the spectrum normalization.
The spectrum shape is largely controlled by the spots' scaling
size $p_s$ alone. In general, the spectra are slightly tilted
(and we numerically compute the dependence of the
tilt on $p_s$), but with an
extra suppression of power at low $l$.
In Sec.\ref{2cor} we derive expressions for
the two-point correlation function $C(\theta)$
and its cosmic variance. Although closer to experiment,
$C(\theta)$ appears to be a very bad discriminator
between texture and inflationary scenarios.
The most original result proved in this paper is presented
in Sec.\ref{ltx}, where we ask how many textures are responsible
for a given $a^l_m$. For high $l$, this number goes like $l$, suggesting
Gaussian behaviour. For low $l$ we find that the $a^l_m$ are
mostly due to the effects of a single texture: the last
(that is, the closest) texture. This ought to impart a
peculiar non-Gaussian signal to the low $a^l_m$, a fact which
we  prove explicitly. The purpose of the sequel
to this paper \cite{ltx2} is to devise a low-$l$ data-analysis
technique (in the moulds of \cite{conf}) capable of fully characterising
this effect.

\section{Analytical spot-throwing}\label{ITT}
In topological defect scenarios the CMBR fluctuations
at large angles are due only to the
ISW (integrated Sachs-Wolfe) effect caused by the defect
network (\cite{txpen}).
For a standard recombination scenario
the validity range for this approximation
is $2\leq l\leq 30$. If early reionization occurs, the
upper limit for this range becomes the value of $l$
corresponding to the angular scale of the horizon
at  ``last scattering'' (defined in
some conservative way). As it is not clear which
is the most sensible scenario we take as a working
hypothesis the choice which maximizes the
applicability of our model: standard recombination.
Now if we assume that the statistical
scaling of the texture fields extends to their associated
time-dependent metric perturbations, then we may expect
the ISW spots themselves to scale. This allows the direct
modelling of the statistics for texture induced spots
as they appear in the sky. From these, we can then write
an expression for the $a^l_m$ joint distribution function
for $2 \leq l\leq 30$. For simplicity, we will assume that
the texture spots are independent (not required by the
formalism, but a good approximation). Then each texture
angular position in the sky is uniformly distributed.
Furthermore its time position in the sky cone can be
associated with a modified
Poisson process in time, in which the
probability of an event is allowed to
vary. Let $n$ be the average number of
texture spot producing configurations
per horizon volume. Then their average volume density is
\begin{equation}
\rho={2n\over 9\pi t^3}
\end{equation}
and using the relation between
the proper area $dS$, the time of emission
$t$, and  the apparent solid angle $d\Omega$:
\begin {equation}
dS=d\Omega{\left(3t{\left({\left(t_0\over t\right)}^{1/3}-1\right)}
\right)}^2
\end{equation}
we find that the angular density of spots in the sky per unit
of emission time is
\begin {equation}
\delta(t)={2n\over \pi t}
{\left({\left(t_0\over t\right)}^{1/3}-1\right)}^2\, .
\end{equation}
It is convenient to introduce the variable $y=\log_2 (t_0/t)$,
the expansion time number as we go backwards in time. Each
$\Delta y=1$ represents one expansion time, starting from $y_0=0$
(here and now) and going as far as  $y_{ls}=\log_2(t_0/t_{ls})$,
when we hit the
last scattering surface. In the standard recombination scenario
$y_{ls}\approx 15$. In terms of $y$ the spot density in the
sky is
\begin {equation}
N(y)=2n\log(2) {\left(2^{y/3}-1\right)}^2\;.
\end{equation}
Let us now introduce a Poisson-type process in $y$, with a
variable probability density $N(y)$. If $p_n(y)$ is the
probability of having exactly $n$ textures at $y'<y$, then
\begin {equation}\label{hieq}
{dp_n\over dy}=N(y)(p_{n-1}-p_n)\quad ,\quad p_{-1}=0\; .
\end{equation}
This hierarchy of equations fully specifies the statistics
of the textures' time-position in the sky.
{}From the $p_n(y)$ one can also derive the probability density of the
$n^{th}$ texture position $y_n$ by means of
\begin {equation}\label{pyn}
P_n(y_n)=p_{n-1}(y_n)N(y_n)\; .
\end{equation}
However one should be wary of the $y_n$ variables, as they are
not independent. Not only are their ranges constrained by
$y_1<y_2<...<y_n$ but also their statistical dependence is
patent in the conditional distribution functions.
{}From ($\ref{hieq}$) and ($\ref{pyn}$) one has that
\begin {equation}
P_1(y_1)=N(y_1)e^{-\int_0^{y_1}dy\, N(y)}
\end{equation}
but then
\begin {equation}
P_2(y_2|y_1)=N(y_2)e^{-\int_{y_1}^{y_2}dy\, N(y)}\;
\end{equation}
which is dependent on $y_1$. From these expressions we can
derive the joint distribution function
\begin {equation}
P_{12}(y_1,y_2)=N(y_1)N(y_2)e^{-\int_0^{y_2}dy\, N(y)}
\end{equation}
which misleadingly factorizes. More generally one has
\begin {equation}\label{pjoint}
P_{1...n}(y_1,y_2,...,y_n)=N(y_1)N(y_2)...N(y_n)
e^{-\int_0^{y_n}dy\, N(y)}\; .
\end{equation}
The statistics contained in $P_n$ or $p_n$ refer to a sequence
of textures ordered in $y$. Ordering will be crucial in Section
\ref{ltx} when we uncover the last texture. However for some
other purposes (like the derivations in Section \ref{aps})
we may simply consider unordered textures. Then
we have a total of $N_t =\int^{y_{ls}}_0dy\, N(y) \gg1$
textures in the sky. Their positions $y_n$ are independent
random variables with a distribution
\begin{equation}\label{stat2}
P_n(y_n)={N(y_n)\over N_t}
\end{equation}
for all $n$, which is considerably simpler.

We now model the texture spot patterns. These ought to follow some sort
of statistical scaling law, where the scale angle is
\begin {equation}\label{om}
\theta^s(y)=
\arcsin{\left(\min{\left( 1,{p_s\over2^{y/3}-1}\right)}\right)}
\end{equation}
for a pattern laid at time $y$. Here $p_s$ is the scaled impact parameter
($p_s=p/(3t)$) giving $\sigma_s=\pi p_s^2$, the scaled cross section for
photon anisotropies. Causality requires that $p_s<1/2$. It may happen
that $p_s$ is a random variable itself,
but we shall ignore this complication.
For small, nearly circular spots with profile $W^s(\theta,y)$, scaling
implies that
\begin {equation}\label{sc}
W^s(\theta,y)={\overline W}(x)
\end{equation}
with $x=\theta/ \theta^s$.
This is a sensible scaling  law even for $y<1$, although for $y<1$
the exact scaling law should be more complicated, as the photons
propagating through the scaling metric no longer move along parallel
trajectories. In particular, if $y<y_0=3\log(1+p_s)$, we
live inside a texture and are enclosed by its pattern.
Unless $(2np_s)^3\approx 1$ textures like these are very rare,
and the approximation ($\ref{sc}$)
should produce good enough final results.

We seek to write the $a^l_m$ coefficients for skies filled with nearly
circular spots described by the above statistics. For each
of these spots let us first point the $z$-axis at their centre,
and perform the spherical harmonic decomposition in that frame.
We obtain an expression of the form
\begin {equation}\label{almn}
{\tilde a}^l_m(n)= a_n W^{sl}(y_n).{\sqrt {2l+1\over 4\pi}}.\delta_{m0}
+\epsilon^l_m
\end{equation}
where $W^{sl}$ is the Legendre transform of the scaled profile:
\begin {equation}\label{omsl}
W^{sl}=2\pi\int^1_{-1} dz\, P^l(z) W^s(z,y_n)
\end{equation}
and $\epsilon^l_m$ is a perturbation induced by the eccentricity.
This perturbation is negligible,
even if the spots are not very nearly circular,
but $l\ll[1/\theta^s(y_n)]$. The extra factor $a_n$ in (\ref{omsl})
accounts for the texture brightness, and is a random variable
with a $y_n$-independent distribution function. Performing a
rotation to a general frame, where the $n^{th}$ texture
has coordinates $\Omega_n$, and summing over all textures one finally
obtains:
\begin{equation}\label{alms}
a^l_m=\sum_n a_n W^{sl}(y_n)Y^{l\star}_m (\Omega_n)\; .
\end{equation}
The variables $a_n$ are independent but equally distributed.
Their distribution has to be determined from the simulations.
The $\Omega_n$ are uniformly distributed, and from them one can
determine the $Y^l_m(\Omega_n)$ distributions
(but with care, as the $Y^l_m$ are not independent).
Finally from ($\ref{pjoint}$), ($\ref{om}$), and ($\ref{omsl}$)
one can determine the $W^{sl}_n$
distributions. The $W^{sl}_n$ makes the various terms in ($\ref{alms}$)
dependent variables, rendering an analytic approach for the
$a^l_m$ distributions through ($\ref{alms}$) unfeasible. In any case
finding the marginal distributions of the $a^l_m$
would not be the end of the story, since the $a^l_m$
are necessarily dependent random variables (see \cite{conf}).
Hence we still would have to find the joint distribution
$F(a^{l_1}_{m_1},a^{l_2}_{m_2},...)$ for a complete
solution to the problem of cosmic variance in texture scenarios.
Expression ($\ref{alms}$) constitutes a formal solution to this problem
which, despite all the obstacles to an analytical solution,
is particularly
suited to  Monte Carlo simulations. It separates the individual
effect of each texture on each of the $a^l_m$. It also
factorizes each texture contribution into three factors.
The $Y^{l\star}_m$ are purely geometrical factors which should be present
in any SSS (statistically spherically symmetric) theory
(see \cite{conf} for a definition).
The $W^{sl}_n$ are a reflection of the structure
of the sky (foliation into expansion times and horizon volumes)
and should be present whenever there is scaling. Finally,
the $a_n$'s factor out all that is peculiar to the texture model under
consideration, and will have to be measured from detailed texture
simulations.

\section{The angular power spectrum}\label{aps}
\subsection{An expression for the $C^l$}\label{apscl}
Throughout this paper we will use the notation $C_l$
for the angular power spectrum of realizations, and $C^l$
for its ensemble average. Whereas $C_l$ is a random variable,
$C^l$ is a number. In any SSS theory one has
\begin{equation}\label{cl0}
<a^l_m a^{l'\star}_{m'}>_{obs}=C^l\delta^l_{l'}\delta^m_{m'}\; .
\end{equation}
The $C^l$ can be estimated from the observable
\begin{equation}\label{cl1}
C_l={1\over 2l+1}{\sum_{m=-l}^l}|a^l_m|^2
\end{equation}
since $<C_l>_{obs}=C^l$ in any SSS theory. Cosmic variance introduces
error bars of size $\sigma(C_l)~$ into this estimate. We start by
deriving an exact expression for the $C^l$.
{}From ($\ref{alms}$) and ($\ref{cl1}$) we have that
\begin{equation}
C^l=<C_l>={1\over 2l+1}{\sum _m}{\sum_{n n'}}
<a_n a_{n'}><W^l_n W^l_{n'}><Y^{l\star}_m(\Omega_n) Y^l_m(\Omega_{n'})>
\end{equation}
but since
\begin{equation}
<Y^{l\star}_m(\Omega_n) Y^l_m(\Omega_{n'})>_{obs}={\delta_{nn'}\over 4\pi}
\end{equation}
we have simply
\begin{equation}\label{clsum}
C^l={<a^2>\over 4\pi}{\sum_n}<W^{l\, 2}_n>\; .
\end{equation}
Computing the averages using the unordered statistical description
($\ref{stat2}$) we can write
\begin{equation}\label{clyf}
C^l={<a^2>\over 4\pi}{\int^{y_{ls}}_0} dy\, N(y) W^{ls2}(y)\; .
\end{equation}
If $p_s$ is a random variable with distribution $f(p_s)$,
then expression (\ref{clyf}) still holds but now with
$W^{sl2}$ replaced by $\int dp_sf(p_s)W^{sl2}(y,p_s)$.
Expression (\ref{clyf}) lends itself to be written as
\begin{equation}\label{cltot}
C^l={\int^{y_{ls}}_0} dy\, C^l(y)
\end{equation}
thus defining a density
\begin{equation}\label{cldens}
C^l(y)={<a^2>\over 4\pi}N(y) W^{ls2}(y)
\end{equation}
measuring the contribution of textures
living at time $y$ to a given $C^l$.
\begin{figure}
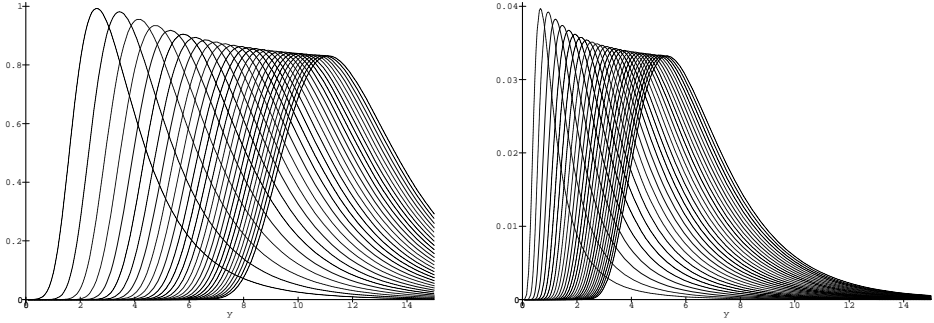

  \begin{center}
    \leavevmode
    {\hbox %
   {\epsfxsize = 6cm
    \epsffile{cly1.ps}}{\epsfxsize=6cm \epsffile{cly2.ps} }}
  \end{center}
\caption{The density $C^l(y)$ multiplied by $l(l+1)$ for $l$ from 2
to 30  for $(n=1,p_s=0.4)$ and $(n=1,p_s=0.08)$
textures with Gaussian profile spots.}
\label{cly}
\end{figure}
In Fig. \ref{cly} we have plotted $l(l+1)C^l(y)$ for $(n= 1,p_s=0.4)$
and $(n=1, p_s=0.08)$ textures with Gaussian profile spots,
for $l$ from 2 to 30. These
plots are to be confronted with the common belief that a given
$C^l$ is mostly due to textures living at a time when the apparent
size of the horizon equals the multipole angular scale.
We find instead that $C^l(y)$ peaks for $y$: $ \Omega^s(y)\approx
{4\pi\over l(l+1)}$,
that is, when the textures' apparent size fits
the multipole angular scale.
Furthermore the peaks comprise several generations and an
accurate prediction for the $C^l$ spectrum can never neglect any
of the integrand in ($\ref{cltot}$).
Changing $n$ only affects the normalization of
$C^l(y)$. Reducing $p_s$ not only decreases the
normalization but also shifts the peaks to
the left necessarily affecting the shape of the
integrated $C^l$. In general as $l$ increases
the peak's heights (multiplied by $l(l+1)$) are slightly reduced,
but their width increases. The delicate balance of these two effects
will determine whether the $C^l$ spectrum is flat or not.

\subsection{Profiles and  reduced $a$ and $p_s$}\label{apsprof}
It is not clear what profile ${\overline W}$ real
texture spots have.
Simulations \cite{txpen,borr}  suggest that the well-known SSSS
profile  may be oversimplified.
We find that the exact profile form has little impact on the $C^l$-spectra.
For intermediate $l$, profiles do not matter at all, and for low $l$
they induce differences easily confused by cosmic variance.
The issue of matching free parameters for different profiles is, however,
non-trivial. We have considered the following profiles:

1.-{\bf Hat profile}. Defined by
\begin{equation}
{\overline W}^s(x)=H(1-x)
\end{equation}
where $H$ is the Heaviside function. Its Legendre transform is
\begin {equation}\label{Omsl}
W^{sl}=2\pi\int^1_{\cos{\theta^s}} dz\, P^l(z)
={2\pi\over 2l+1}{\left( P_{l+1}(\cos{\theta^s})-P_{l-1}
(\cos{\theta^s})\right)}\; ,
\end{equation}
the $l$-weighted scaled solid angle.

2.-{\bf Toy hat profile}. We can skip the complications induced by
the Legendre polynomials approximating the hat profile by
\begin{equation}
W^{sl2}= \left\{ \begin{array}{ll}
		\Omega^{s2}&\mbox{for }\Omega^s <  \Omega^{l(sat)}
					={4\pi\over l(l+1)}\\
		\Omega^{l(sat)2}&\mbox{for }\Omega^s >\Omega^{l(sat)}
		     \end{array}
	\right.
\end{equation}
with
\begin{equation}
\Omega^s=2\pi(1-\cos{\theta^s})\; .
\end{equation}
the scaled solid angle.

3.-{\bf Gaussian profile}. A large angle generalization of
the well-known small
angle Gaussian window:
\begin{equation}
W^{ls}=\Omega^s e^{-{(l+{1\over 2})^2\theta^{s2}\over 2}}\; .
\end{equation}

4.-{\bf SSSS profile}. A profile of the form (e.g. \cite{rd})
\begin{equation}
W^s(\theta,y)\propto {\theta^s\over {\sqrt {2\theta^{s2}+\theta^2}}}
\end{equation}
with some sort of cut off at
\begin{equation}
\theta^h(y)=\arcsin{\left(\min{\left( 1,{{1/2}\over2^{y/3}-1}\right)}\right)}
\end{equation}
the angle subtended by the horizon. We define this cut off with the Legendre
transform:
\begin{equation}
W^{sl}= \left\{ \begin{array}{ll}
		2\pi \theta^h \theta^{s} e^{-{\sqrt 2}\theta^s/\theta^h}
		&\mbox{for } l\leq[1/\theta^h]\\
		2\pi\theta^s {e^{-{\sqrt 2}\theta^s l}\over l}&\mbox{for }
		l\geq[1/\theta^h]\; .
		     \end{array}
	\right.
\end{equation}

The hat profile is suggested by the
belief that ${\delta T\over T}$ circular averages are all that
matters for $2\leq l\leq 30$. Circular averages have been extensively studied
(eg. \cite{vir}) and some useful properties have been found
for them.
The toy hat profile, due to its simplicity,  is ideal for an heuristic
explanation of the results obtained (Sec.\ref{hrt}).
The Gaussian profile is particularly amenable to analytical work,
thus providing a good check on the numerics. Furthermore it
is probably a good approximation to the peaks found in \cite{borr}.
Finally, there is no harm in considering the SSSS profile, for
it might have something to do with reality after all.

With the ${\overline W}$ defined as they are, the functions
$C^l(y)$
peak for different values of $y$. To facilitate
comparison we should therefore define a reduced $p_s$:
\begin{equation}\label{ovp}
{\overline p_s}={p^{hat}\over 1.8}={p^{toy}\over 2}=p^{gauss}
\end{equation}
which ensures that, for high $l$, $C^l(y)$ peaks at the same
$y_{max}$ for the same ${\overline p_s}$ for all profiles.
For the SSSS profile $C^l(y)$ peaks always for $y:$ $\theta^h(y)\approx {1
\over l}$. This is understandable as the integrated anisotropy
increases logarithmically
as we go away from the core. As a
result, the effective cross-section for SSSS textures is the horizon
area, for all $p_s$. For this reason we will always treat the SSSS
profile separately, as it does not really have a parameter $p_s$.

It also happens that
the value of $C^l(y_{max})$ depends on the profile, taking the form
\begin{equation}\label{clmax}
C^l(y_{max})={<a^2>.2n\log(2).\pi p_s^2\over l(l+1)}.\alpha
\end{equation}
with $\alpha^{hat}\approx 0.58$, $\alpha^{toy}=1$,
$\alpha^{gauss}=1/(4e)$.
%$\alpha^{SSSS}=e^{-{\sqrt{2}}}$.
By defining
a reduced $a$ as
\begin{equation}\label{ova}
{\overline a}=a{\sqrt \alpha}{p_s\over {\overline p_s}} =
1.05 a^{hat}=2a^{toy}={a^{gauss}\over 2{\sqrt{e }}}
\end{equation}
we can ensure that $C^l(y_{max})$ for high $l$
is the same for all the profiles
with the same $<{\overline a}^2>$ and ${\overline p_s}$.

\begin{figure}[t]
  \begin{center}
    \leavevmode
    \epsfxsize = 6cm
    \epsfysize = 6cm
    \epsffile{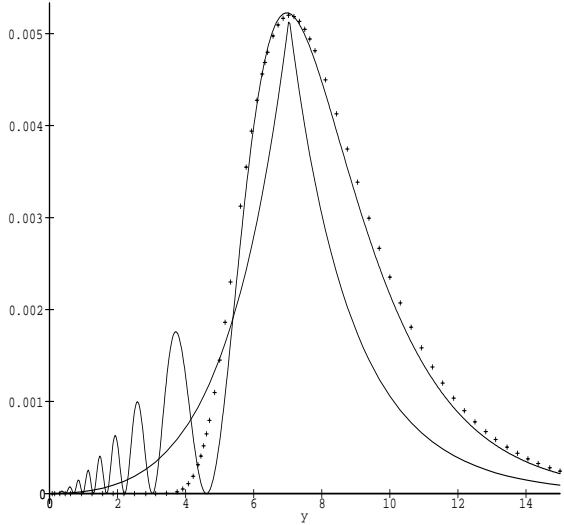}
  \end{center}
  \caption{$C^{20}(y)$ for the hat profile (oscillating line),
 the toy hat profile (non-oscillating line), and Gaussian profile
(with points). For all the curves ${\overline p}_s=0.2$ and $<{\overline
a}^2>$ is the same.}
  \label{vprof}
\end{figure}

In Fig. \ref{vprof} we have plotted $C^{20}(y)$ in units of
$<{\overline a}^2>n$, with
${\overline p_s}=0.2$, for the first three profiles
proposed. We notice that for $y>y_{max}$ all the $C^l(y)$
have the same form, as for all profiles $W^{ls}\propto\Omega^s$
(even though the proportionality constants may be different).
Consequently, in this region $C^l(y)\propto\Omega^s (y)\propto1/N(y)$
regardless of the profile. For $y\ll y_{max}$ we have $W^{ls}\ll
\Omega ^s$ and so $C^l(y)$ is proportional to $N(y)$ or smaller than that.
The exact form of $C^l(y)$ in this region
is profile dependent, as it involves details of how
$W^{ls}(\Omega^s)$ starts to cut off at a particular scale.
While this portion of the integrand is not the
main contribution to the total $C^l$ it is clear that the detailed
form of the spectrum is going to be profile dependent.
It remains to be seen if these differences have any meaning
once cosmic variance error bars are taken into account.

It is also in the $y\ll y_{max}$ region that $C^l(y)$ may
be affected by the existence of non-spherical spot modes
($\epsilon^l_m$ in (\ref{almn})). These are the only source of
inaccuracies in the model proposed, and so whatever mistakes
we make will be in this region. Like profile differences,
spots' ellipticity can only affect the final $C^l$ in the fine detail.
Most probably, ellipticity affects $C^l$ even less
than the profile, as the spots seen in the simulations
appear to be very circular.

\subsection{An expression for $\sigma^2(C_l)$}\label{apsscl}
We now turn to the issue of cosmic variance in the $C_l$.
This turns out to depend on the nature of the theory.
It is known that in Gaussian theories the $C_l$ are
$\chi^2_{2l+1}$-distributed with a variance
\begin{equation}\label{cvg}
\sigma^2_G(C_l)=C^{l2}{2\over 2l+1}\; .
\end{equation}
In some of the literature these are also the cosmic variance error bars
attributed to non-Gaussian theories. We will use our model to
compute directly $\sigma^2_{TX}(C_l)$.
{}From ($\ref{alms}$) we have that
\begin{eqnarray}
<C^2_l>_{obs}&=&{1\over (2l+1)^2}{\sum_{mm'}}{\sum_{(n_1n_2n_3n_4)}}
<a_{n_1}a_{n_2}a_{n_3}a_{n_4}><W^l_{n_1}W^l_{n_2}W^l_{n_3}W^l_{n_4}>
\nonumber \\
&&<Y^{l\star}_m(\Omega_{n_1})Y^{l}_m(\Omega_{n_2})
Y^{l\star}_{m'}(\Omega_{n_3})
Y^{l}_{m'}(\Omega_{n_4})>\; .
\end{eqnarray}
The last factor can only be non-zero in one of the following cases:
\begin{itemize}
\item if $n_1=n_2$ and $n_3=n_4$ but
$n_1\neq n_3$, giving rise to $(2l+1)^2$ terms,
\item  if $n_1=n_3$ and
$n_2=n_4$ (but $n_1\neq n_2$) or $n_1=n_4$ and $n_2=n_3$ (but
$n_1\neq n_2$), giving rise to  a total of $2(2l+1)$ terms,
\item if $n_1=n_2=n_3=n_4$, giving rise to a single term.
\end{itemize}
Putting all these terms together one has
\begin{equation}
<C^2_l>_{obs}={<a^2>^2\over (4\pi)^2} {\sum_{n\neq n'}}
<W^{l2}_nW^{l2}_{n'}>{\left( 1+{2\over 2l+1}\right)}
+{<a^4>\over (4\pi)^2}{\sum_n}<W^{l4}_n>\; .
\end{equation}
Subtracting off $C^{l2}$ written as ($\ref{clsum}$),
using (\ref{stat2}), and neglecting terms in $1/N_t$,
one finds after some
algebra
\begin{equation}\label{cov}
\sigma^2_{TX}(C_l)=C^{l2}{\left(
{2\over 2l+1} +{\left(1+{\sigma^2(a^2)\over< a^2>^2}\right)}
{{\int^{y_{ls}}_0} dy\, N(y) W^{ls4}(y)\over
({\int^{y_{ls}}_0} dy\, N(y) W^{ls2}(y))^2}\right)}\; .
\end{equation}
\begin{figure}[bt]
  \begin{center}
    \leavevmode
    \epsfxsize = 6cm
    \epsfysize = 6cm
    \epsffile{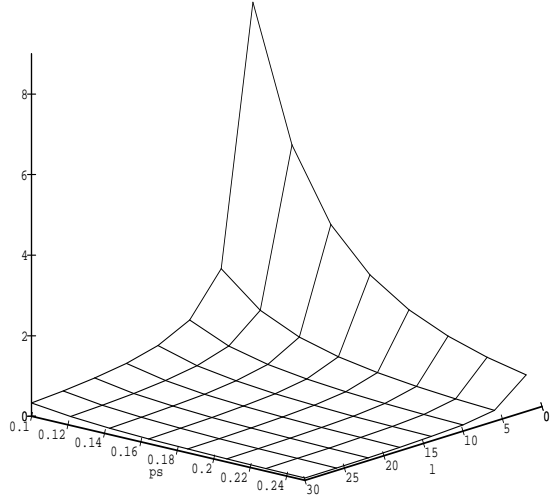}
  \end{center}
  \caption{The relative non-Gaussian
excess variance ${\tilde\epsilon}^l(p_s)$ for Gaussian profile spots. }
  \label{cvcorr}
\end{figure}
We note that the cosmic variance in the $C^l$ in texture scenarios
is always larger than in Gaussian theories. We define $\epsilon^l$,
the relative size of the excess variance, as
\begin{equation}
\epsilon ^l={\sigma^2_{TX}(C_l)-\sigma^2_G(C_l)\over \sigma^2_G(C_l)}\; .
\end{equation}
$\epsilon^l$ is a function of $n$, $p_s$, and the profile. In general
$\epsilon^l(n,p_s)=\epsilon^l(1,p_s)/n$, so we define ${\tilde\epsilon}
^l(p_s)=\epsilon^l(1,p_s)$, the excess variance per unit of $1/n$.
The quantity $\epsilon^l$ can be seen as a (theoretical)
indicator of how non-Gaussian a given multipole is. A texture
theory with $n\rightarrow\infty$ will be a Gaussian theory.
A low $n$ theory will be very non-Gaussian, at least for
some $l$. In Fig. \ref{cvcorr} we have plotted ${\tilde\epsilon}
^l(p_s)$ for spots with a Gaussian profile. Generally the
excess variance is very small for high $l<30$. For low $l$
(say from 2 to 5) the correction can be considerable,
the more so the smaller the $p_s$.

\subsection{$C^{l\star}$-spectra for
different ${\overline p}_s$ and profiles}
\label{apsclres}
\begin{figure}[tb]
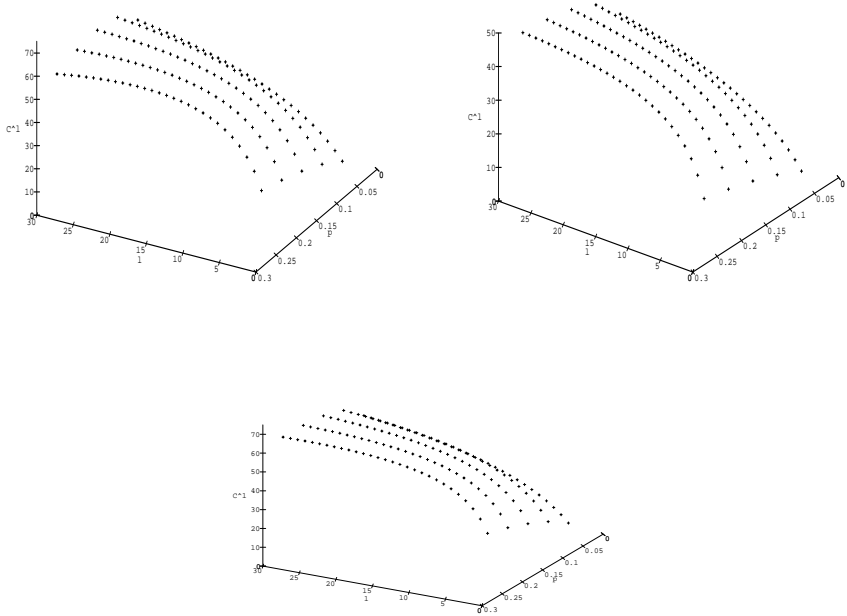

  \begin{center}
    \leavevmode
	{\hbox %
   { \epsfysize =6cm\epsfxsize = 6cm
    \epsffile{rclh.ps}}{ \epsfxsize = 6cm\epsfysize=6cm
    \epsffile{rclth.ps}}
	}
{\epsfxsize=6cm \epsffile{rclgs.ps}}
  \end{center}
\caption{$C^{l\star}$ spectra for the hat, toy hat, and Gaussian profiles.}
\label{clh}
\end{figure}

{}From ($\ref{cldens}$) we see that the parameters
$<a^2>$ and $n$ can only affect the overall normalization
of the spectrum. Also (\ref{clmax})
suggests that the ``average'' spectrum normalization
is proportional to $\pi p_s^2$, although $p_s~$
may affect the spectrum shape as well. Furthermore a roughly
scale invariant ($\propto 1/l^2$) spectrum can be expected.
We do want more than a rough prediction, though, so we define a
reduced spectrum
\begin{equation}\label{cstar}
C^{l\star}={4\pi l(l+1)C^l\over {<{\overline a}^2>}.2n\log(2).\pi
{\overline p}
_s^2}\; .
\end{equation}
$C^{l\star}$ describes in
detail departures from scale invariance and factors out
all that contributes only to the normalization of the spectrum.
It depends only on the profile and ${\overline p}_s$.
In Fig. \ref{clh}  we have plotted the
$C^{l\star}$ spectrum for various values of ${\overline p}_s\leq1/4$
(to ensure causality), for the hat, toy hat, and Gaussian profiles.
The spectrum normalization still depends on the profile. Even though
the peak heights in $C^l(y)$ have been matched by our definitions of
${\overline p}_s$ and ${\overline a }$, the differences in the peak's
shapes and in $C^l(y)$ away from the peak produce different
integrated $C^l$ for different profiles. Apart
from this it is remarkable that the $C^{l\star}({\overline p}_s)$
spectra shapes
do not depend on the profile in any significant way.
For intermediate scales ($l_1<l<30$, for some
$l_1$), the spectra are
scale-invariant if ${\overline p}_s$ is not too small. If
${\overline p}_s$ is small, the spectrum is slightly tilted.
Whatever the case, for low $l$ ($l<l_1$) the power is suppressed
relative to a scale invariant (or slightly tilted) spectrum.
An exact evaluation of the tilt, $l_1$, and low $l$ suppression
factor can only be done numerically. An approximate
formula  and heuristic explanation of this effect is given in
Sec.\ref{hrt}.

\begin{figure}[tb]
  \begin{center}
    \leavevmode
	{\hbox %
{\epsfxsize=6cm \epsffile{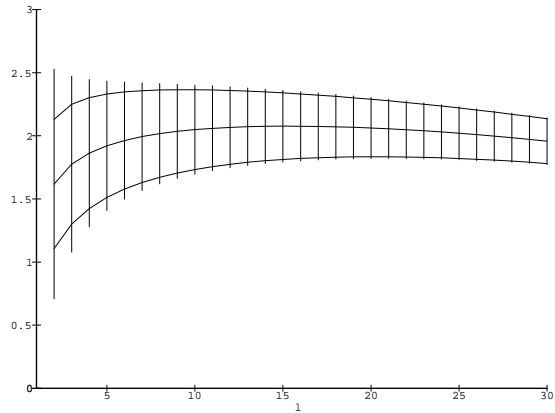}} }
  \end{center}
\caption{$C^{l\star}$ spectrum (middle line) and its Gaussian (lines)
and non-Gaussian (bars) cosmic variance
error bars for spots with a  SSSS profile with $n$=0.25 and $p_s=0.2$.}
\label{clssss}
\end{figure}
\subsubsection{SSSS collapses}
We have applied our formalism to concrete choices of ($n,
{\overline p}_s$) and profile which have often been
suggested by simulations. In \cite{durr} there is the suggestion
that it is a good approximation to consider that only
SSSS collapses followed by unwindings cause CMBR spots.
This leads to a scenario with a low spot number density
($n\approx 0.25$), and regardless of $p_s$ (which is a
random variable in this case), to an effective spot
size of the order of the horizon angular size. We have pointed out
before that
$n$ and $p_s$ do not matter for the spectrum shape
produced by SSSS spots. We have plotted in Figure \ref{clssss}
the $C^{l\star}$ spectrum (${\overline p}_s=p_s$,
${\overline a}=a$, no division by $\pi p_s^2$ in ($\ref{cstar}$))
for SSSS spots with $p_s=0.2$
(for definiteness), with cosmic variance error bars
computed from ($\ref{cov}$). A flat spectrum without any significant
low $l$ cut-off is obtained, in agreement with
\cite{txpen,durr}. This is a general feature
for any profile whenever ${\overline p}_s$ is sufficiently
large. The only novelty is the large non-Gaussian
correction to the cosmic variance
in this scenario, due to the low value of $n$
(recent reruns of \cite{txpen} have shown an abnormally
large cosmic variance).

\begin{figure}[t]
  \begin{center}
    \leavevmode
	{\hbox %
{\epsfxsize=6cm \epsffile{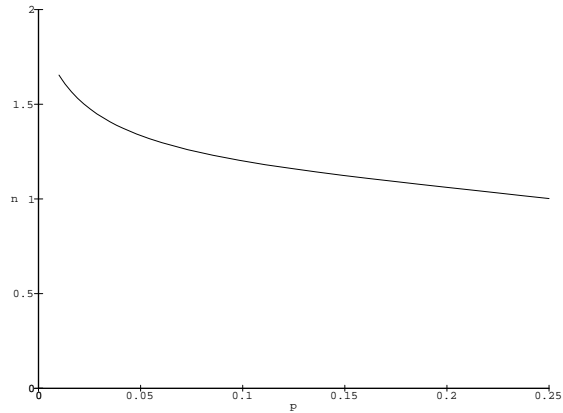}}
}
  \end{center}
\caption{The spectral index $n^i$ for high $l$ as a function of
${\overline p_s}$.}
\label{index}
\end{figure}
\subsubsection{Gaussian spots}
\begin{figure}[t]
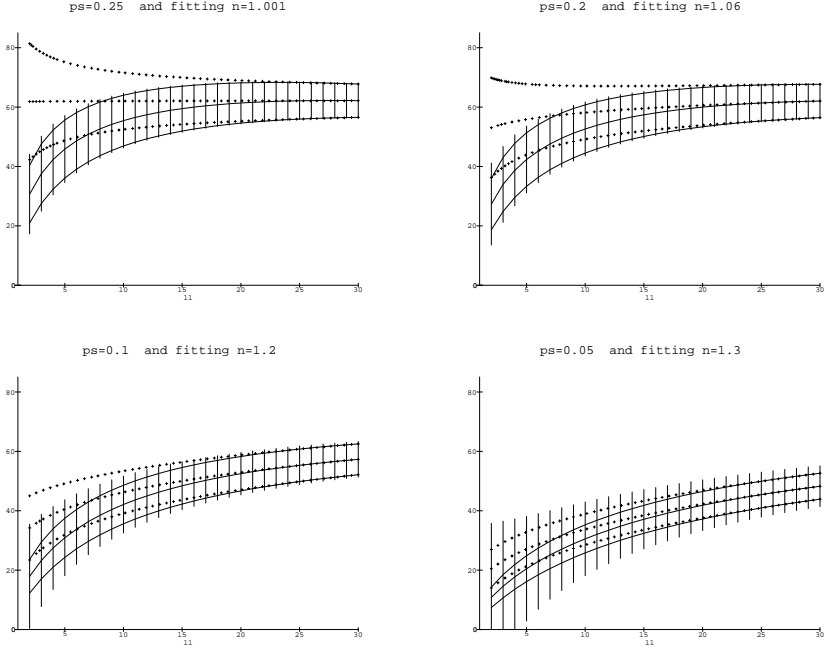

  \begin{center}
    \leavevmode
	{\hbox %
{\epsfxsize=6cm \epsffile{clvg0.ps}}
{\epsfxsize=6cm \epsffile{clvg1.ps}}
}
{\hbox %
{\epsfxsize=6cm \epsffile{clvg2.ps}}
{\epsfxsize=6cm \epsffile{clvg3.ps}}
}
  \end{center}
\caption{$C^{l\star}$ spectrum (middle line) with its
Gaussian (lines) and full (bars)  cosmic variance error bars
for Gaussian profile spots with ${\overline p}_s=0.25$,  $0.2$,
$0.1$,  and 0.05,
confronted with fitting Gaussian tilted
spectra with the same normalization (points).}
\label{clbr1}
\end{figure}

In \cite{borr}, on the other hand, it was found that both realistic
(non-symmetric) unwindings and concentrations of gradient
energy produce CMBR spots. Together, these two types of effect
bring the spot number density up to $n\approx 1$. The spots
produced by the two effects appear to be indistinguishable.
Their profile is clearly non-SSSS, and is better
approximated by a Gaussian profile with a small ${\overline p}_s$,
around $0.1$. A clear evaluation of ${\overline p}_s$ has not
been provided by the simulations. These simulations
are essentially flat space-time
simulations (\cite{borr}) in which the horizon
size is ambiguously defined.
Hence we leave ${\overline p}_s$
as a free parameter. Low ${\overline p}_s$
produce slightly tilted spectra even for intermediate $l$
(cf. Fig. \ref{clh}). A natural question is
how the tilt depends on ${\overline p}_s$. We fitted the $C^{l\star}$
spectra for Gaussian profiles to the inflationary type of spectra
\begin{equation}
C^l\propto{\Gamma {\left(l+{n^i-1\over 2}\right)}
\Gamma{\left(9-n^i\over 2 \right)}\over \Gamma{\left(l+{5-n^i\over 2}\right)}
\Gamma {\left(3+n^i\over 2\right)} }
\end{equation}
where $n^i$ is the spectral index. The fit was performed for
$l\in(25,30)$, and the resulting $n^i=n^i({\overline p}_s)$
function is plotted in
Fig. \ref{index}. For ${\overline p}_s>0.15$ we find $n^i\approx 1$,
but for ${\overline p}_s=0.05$ and ${\overline p}_s=0.1$ we have
respectively $n^i=1.3$ and $n^i=1.2$, for instance.
Hence the importance of an accurate measurement for ${\overline p}_s$
as it will provide the intermediate scale spectral index for textures.
Still, this is not the end of the story. In Fig. \ref{clbr1} we
plotted the $C^{l\star}$ spectrum for four values of ${\overline p}_s$
with 1-$\sigma$ cosmic variance error bars. Superposed on them
are the fitting inflationary type of spectra, with their
Gaussian cosmic variance error bars. If $p_s$ is not too
small, a 1-$\sigma$ differentiation
between the two theories arises for $l=2$ and $3$ although the cosmic
confusion at intermediate $l$ is almost 1. This is somewhat
surprising, as the cosmic variance in the low $C^l$
is very high (and even higher in texture scenarios). In
spite of this, the extra suppression of low $l$ in texture
scenarios is strong enough to survive cosmic confusion.
It is tempting to connect this effect with the abnormally low
$C_l$ for $l=2,3$ observed by COBE.
However, the experimental error bars make it
unwise to draw any conclusion. This effect also shows
how a concept like the spectral index $n^i$, coined
for inflation, becomes inadequate as a texture spectra qualifier.
Texture spectra have a non-uniform tilt.
If $p_s$ is very small the extra suppression,
although meaningful if the cosmic variance were Gaussian
(high $n$), is completely drowned by the non-Gaussian excess
of cosmic variance. Although somewhat disappointing, this
result also signals strong non-Gaussianity for low-$l$.
At the same time as it renders the $C^l$ spectrum useless,
it suggests that a particularly strong signal ought to exist
in quantities measuring non-Gaussianity.

\subsection{An heuristic interpretation and an approximate formula}
\label{hrt}
The expansion into spherical harmonics acts as a scale filter.
Each harmonic selects spots with an angular scale $\Omega ^s$
smaller than $\Omega^{l(sat)}={4\pi\over l(l+1)}$.
Whatever the spot profile,
if $\Omega^s\ll \Omega^{l(sat)}$ then $W^{ls}\approx \Omega^s$,
but if $\Omega^s\gg \Omega^{l(sat)}$ then $W^{ls}\ll\Omega^s$.
In the latter case, depending on the profile, $W^{ls}$ may either reach a
plateau ($\approx \Omega^{l(sat)}$) or simply decrease
monotonically with $\Omega ^s$. We can understand the shape of
$C^l(y)$ by taking this into account. For $y: \Omega^s <\Omega^{l(sat)}$,
each texture contributes to $C^l$ like $\Omega^{s2}$, but when
we add them up in the $a^l_m$ they may interfere either
constructively or destructively. As a result, only a r.m.s.
fluctuation proportional to ${\sqrt{N}}$ contributes to ${\sqrt{C^l}}$.
The overall contribution from  a given $y$ is therefore proportional to
$N\Omega^{s2}$, which decreases as we go up in $y$. As $y$ increases
we do have more textures in the sky, but their apparent size is also
much smaller, and since $N\Omega\approx const$, the balance makes
$C^l(y) $ decrease. For $y:\Omega^s>\Omega^{l(sat)}$, the filter
starts to act, collecting in a given $C^l$ a contribution
from  each individual texture proportional to $W^{ls2}\ll\Omega^{s2}$.
This may either saturate or cut-off. Whatever the case the contribution
from a given $y$ will be proportional to $N$ or less than that.
The contribution from each texture is smaller than a constant,
and $C^l$ increases as we go up in $y$, simply because there
are more textures creating a r.m.s. fluctuation. Hence there
will be a peak in $C^l(y)$ at $y:\Omega^s(y)\approx\Omega^{l(sat)}$,
and we may expect that most of the contribution to a given $C^l$
will come from the scales where the filter starts to act.
The contribution from these textures is of the order $N\Omega^{l(sat)2}$,
and recalling that $N\Omega^s\approx const$, we can expect a value
of $C^l$ proportional to $\Omega^{l(sat)}\propto 1/l^2$.
This explains why we can always expect  a roughly scale-invariant
spectrum in the  texture scenario. The only exception to this argument
is the case where the $\Omega^s>\Omega^{l(sat)}$ regime does not
exist (or is irrelevant), simply because $\Omega^{l(sat)}$
is greater than $<\Omega_1>$, the average angular size of the last texture.
This is the case for
low $l$, which do not act as a filter for any of the
texture spots, because their cut-off scale is above the angular
size of the last (and largest texture).
Then, the largest contribution to $C^l$ comes from the last texture,
this being the case for all $l:\Omega^{l(sat)}\gg<\Omega_1>$.
The last texture contribution to  these $C^l$ is, of course, independent
of $l$, and proportional to $<\Omega_1^2>$. Hence we can
expect a white noise type of spectrum ($C^l\propto const$) in
this $l$ region. In practice, the transition from white noise
to scale invariance happens very quickly. We never observe a
white noise regime, but only a suppression for the low $l$.

\begin{figure}[t]
  \begin{center}
    \leavevmode
	{\hbox %
   { \epsfxsize = 6cm
    \epsffile{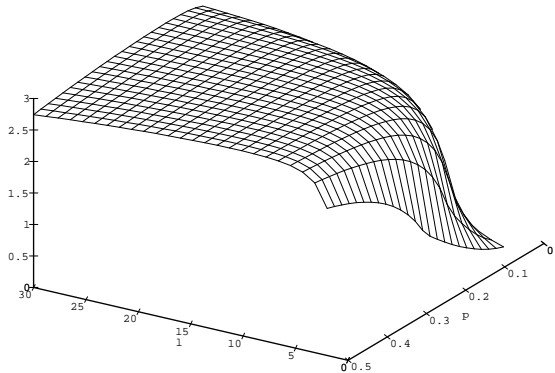}}}
  \end{center}
\caption{Approximate spectra $l(l+1)C^l$ for various values of $p_s$.}
\label{clap}
\end{figure}

This heuristic argument can be converted into an approximate formula
by taking the toy-hat profile and carrying out explicitly
the $C^l$ integration, starting from $y=<y_1>$. Let us define
$l_1$ as
\begin{equation}
\Omega_1= \Omega^s(<y_1>)={4\pi\over l_1(l_1+1)}
\end{equation}
(the $l$-scale of the last texture), and $l_{ls}$ as
\begin{equation}
\Omega^s(y_{ls})={4\pi\over l_{ls}(l_{ls}+1)}
\end{equation}
(the $l$-scale of textures living at
the last-scattering surface). Then
\begin{equation}
C^l\propto {\int^{\Omega_1}_{\Omega_{ls}}}
d\Omega {W^{sl2}\over \Omega^{s2}}
\propto   \left\{ \begin{array}{ll}
		\Omega_1-\Omega_{ls}
		&\mbox{for } l\leq l_1\\
		3\Omega^{l(sat)}-2{\Omega^{l(sat)2}\over \Omega_1}-
		\Omega^{ls}
		&\mbox{for }l\geq l_1
		     \end{array}
	\right.
\end{equation}
giving mathematical expression to what we have said above.
This leads to the approximate spectral formula
\begin{equation}\label{clapp}
C^l\propto \left\{ \begin{array}{ll}
		{1\over l_1(l_1+1)}-{1\over l_{ls}(l_{ls}+1)}
		&\mbox{for } l\leq l_1\\
		{3\over l(l+1)} -{ 2l_1(l_1+1)\over (l(l+1))^2} -
		{1\over l_{ls}(l_{ls}+1)} &\mbox{for }l\geq l_1
		     \end{array}
	\right.
\end{equation}
with $l_1\approx 0.5/{\tilde p}_s$ and $l_{ls}\approx 30/
{\tilde p}_s$ for standard recombination, with ${\tilde p}_s=2
{\overline p}_s$. The approximate spectra were plotted in
Fig. \ref{clap}
with an arbitrary vertical scale.  Confronting this with Fig.
\ref{clh}, we see that
($\ref{clapp}$) does provide a good qualitative description
of the spectra.

\section{The two-point correlation function}\label{2cor}
The two-point correlation function $C(\theta)$ is a tool
sadly closer to experiment than the $C^l$ spectrum
(however see \cite{scott} for a more hopeful point of view).
We will now show that $C(\theta)$ is a very bad discriminator
between inflationary and texture scenarios. $C(\theta)$
is defined as
\begin{equation}
C(\theta)=<{\delta T\over T}(\Omega){\delta T\over T}(\Omega ')>_{obs}
\end{equation}
with $\theta ={\widehat {(\Omega,\Omega')}}$, ($\Omega$ and $\Omega'$
any two fixed directions in the sky). Making use of (\ref{cl0})
this can be written as
\begin{equation}\label{ct0}
C(\theta)={\sum _l}C^l{2l+1\over 4\pi}P_l(\cos\theta)\; .
\end{equation}
The theoretical $C(\theta)$ can be estimated from the sky average of
${\delta T\over T}(\Omega){\delta T\over T}(\Omega ')$
with a fixed $\theta ={\widehat {(\Omega,\Omega')}}$:
\begin{equation}
C_{obs}(\theta)=\int{d\Omega\, d\Omega'\over 8\pi^2}
{\delta T\over T}(\Omega){\delta T\over T}(\Omega ')
\delta({\cos\theta}-{\cos {\widehat {(\Omega,\Omega')}})}
\end{equation}
which after some algebra leads to
\begin{equation}
C_{obs}(\theta)= {\sum _l}C_l{2l+1\over 4\pi}P_l(\cos\theta)
\end{equation}
with $C_l$ defined as in (\ref{cl1}). An expression for $C(\theta)$
in texture scenarios can be found by
replacing (\ref{clyf}) in (\ref{ct0}).
It is curious that a simpler expression can be derived. Inverting
the Legendre transform (\ref{omsl}) one has
\begin{equation}
W^s(z,y)={\sum_l}{2l+1\over 4\pi}W^{sl}(y)P_l(z)\; .
\end{equation}
If we define the Legendre-squared spot profile to be
\begin{equation}
W^{s(2)}(z,y)={\sum_l}{2l+1\over 4\pi}W^{sl2}(y)P_l(z)
\end{equation}
then the correlation function is simply
\begin{equation}\label{ctf2}
C(\theta)={<a^2>\over 4\pi}{\int^{y_{ls}}_0} dy\, N(y) W^{s(2)}
({\cos\theta}, y)\; .
\end{equation}
The profile $W^{s(2)}$ is the Legendre analogue of
$W^s\star W^s$ (where $\star$ is the convolution)
in Fourier analysis. Depending on whether the
profiles are defined in the $\theta$ or in the
Legendre spaces it may be easier to use (\ref{ctf2}) or
to combine
(\ref{ct0}) and (\ref{clyf}).  If one wants to take into account
the telescope beam filtering, one should multiply the
$C^l$ in (\ref{ct0}) by the square of the beam Legendre
transform ${\cal F}^{l2}$. In the numerics we will
use an approximation to the COBE
beam ${\cal F}^l=e^{-{1\over 2} (l+{1\over 2})^2\sigma^2}$
with $\sigma=2\pi (10^o/360^o)$.

\begin{figure}[t]
  \begin{center}
    \leavevmode
	{\hbox %
   { \epsfxsize = 6cm
    \epsffile{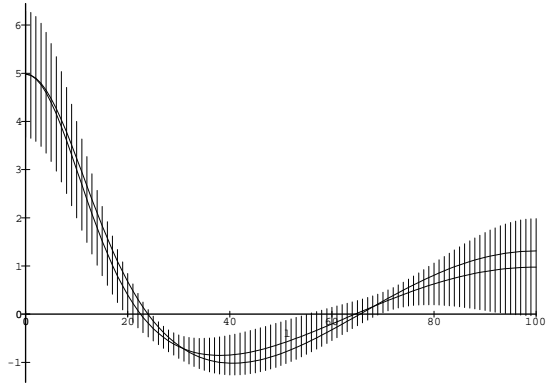}}}
  \end{center}
\caption{$C(\theta)$ for Gaussian spots with ($n=1$, $p_s=0.25$)
with its full cosmic variance error bars, confronted with its fitting
$n=1.001$ theory (top line at $C(\pi)$).}
\label{epsc2}
\end{figure}

We have also derived an expression for the cosmic variance
in $C_{obs}(\theta)$. In Gaussian theories ($\cite{efst}$ with a factor
of 2 correction) we have:
\begin{equation}
\sigma^2_G(C_{obs}(\theta))={\sum_l}\sigma^2_G(C_l)
{\left({2l+1\over 4\pi}\right)}^2 P_l^2
({\cos\theta})
\end{equation}
with $\sigma^2_G(C_l)$ given by (\ref{cvg}). In the texture
scenarios not only does one have to replace $\sigma^2_G(C_l)$
by $\sigma^2_{TX}(C_l)$ (as in (\ref{cov})), but also terms in
$P_lP_{l'}$ (with $l\neq l'$) appear,
reflecting the inter-$l$ correlations. A rather elaborate calculation,
in the moulds of Sec.\ref{apsscl}, brings us to
\begin{equation}\label{sigcorr}
\sigma^2_{TX}(C_{obs}(\theta))={\sum_{ll'}}V_{ll'}C^lC^{l'}
{2l+1\over 4\pi}{2l'+1\over 4\pi}P_l({\cos\theta})P_{l'}({\cos\theta})
\end{equation}
with
\begin{equation}
V_{ll'}={2\delta_{ll'}\over 2l+1}
+{\left(1+{\sigma^2(a^2)\over< a^2>^2}\right)}
{{\int^{y_{ls}}_0} dy\, N(y) W^{ls2}(y) W^{l's^2}(y) \over
({\int^{y_{ls}}_0} dy\, N(y) W^{ls2}(y))
({\int^{y_{ls}}_0} dy\, N(y) W^{l's2}(y))}
\end{equation}
where, again, we have neglected terms in $1/N_t$.
The intensity of the non-Gaussian correction
to the cosmic variance can be measured
by means of the quantity
\begin{equation}\label{epscorr}
\epsilon(\theta)={\sigma^2_{TX}(C_{obs}(\theta))-
\sigma^2_{G}(C_{obs}(\theta))\over
\sigma^2_{G}(C_{obs}(\theta))}
\end{equation}
which depends on $n$, ${\overline p}_s$, and the profile.
The dependence on $n$ is trivial, so we define
${\tilde\epsilon}(p_s,\theta)=\epsilon(n,p_s,\theta).n$.
The numerator in (\ref{epscorr}) can be rewritten as
$<(W^{s(2)})^2>$, showing that the variance in
$C_{obs}(\theta)$ is always larger in texture scenarios than
in similar Gaussian theories. As before, we adopt the attitude
that while this does reduce the predictability of the theory
on $C(\theta)$, it also signals non-Gaussian behaviour
and the need of a non-Gaussian data-analysis approach to
fully make out the predictions of the theory. In particular,
on angular scales where $\epsilon\gg 1$, we know that
it is worth studying the collapsed (to two points) $n$-point
correlation function.
\begin{figure}[t]
  \begin{center}
    \leavevmode
	{\hbox %
   { \epsfxsize = 6cm
    \epsffile{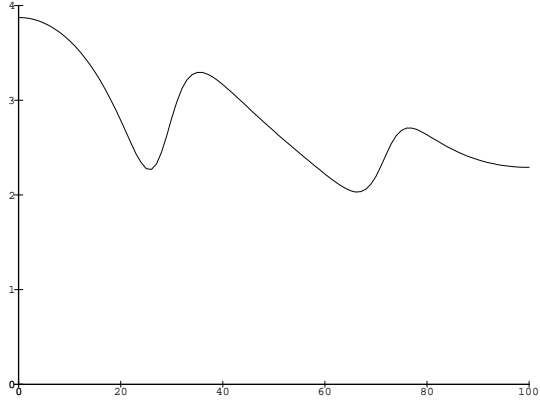}}}
  \end{center}
\caption{$\epsilon(\theta)$ for Gaussian spots with ($n=1$, $p_s=0.25$).}
\label{epsc}
\end{figure}
In Fig. \ref{epsc} we divided $\theta\in(0,\pi)$
into  100 points and plotted
$\epsilon(\theta)$ for Gaussian spots with ($n=1$, $p_s=0.25$).
The fact that the low $l$ multipoles contribute to $C(\theta)$
for all $\theta$ distributes non-Gaussianity nearly
uniformly over $\theta$.
\begin{figure}[t]
  \begin{center}
    \leavevmode
	{\hbox %
   { \epsfxsize = 6cm
    \epsffile{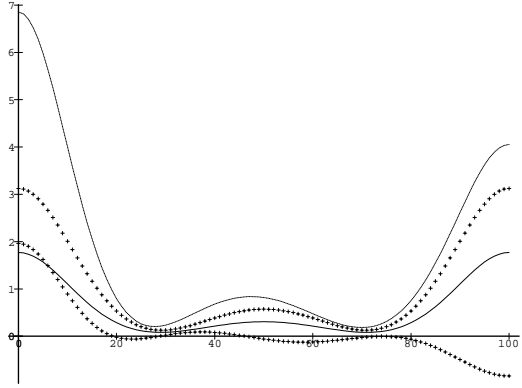}}}
  \end{center}
\caption{$\sigma^2_{TX}$ (top line), and its Gaussian (bottom line),
diagonal (top points), and off-diagonal (bottom points) in
$V_{ll'}$, contributions. }
\label{epsc1}
\end{figure}
In Fig. \ref{epsc1} we analyse the various contributions to
$\sigma^2_{TX}$, separating its Gaussian, diagonal and off-diagonal
components in $V_{ll'} $.
We note that the inter-$l$ correlations responsible for
the off-diagonal elements of $V_{ll'}$ can act so as to
reduce the cosmic variance, an important fact which we shall make use of
in \cite{ltx2}. Naturally, the subtle differences in the
$C^l$-spectra pointed out at the end of Sec.\ref{apsclres}
are completely drowned in $C(\theta)$. As an example
we have plotted in
Fig.~\ref{epsc2} $C(\theta)$ for the above texture theory,
and for its fitting tilted spectrum inflationary theory.
As $C(\theta)$ spreads the low $l$
over all $\theta$, the two theories come out
completely confused. Note however that the normalization
in the $C^l$ and in $C(\theta)$ can be substantially
different (in Fig. \ref{epsc2} we used
a different normalization procedure than in Fig. \ref{clbr1}).

\section{The last texture and non-Gaussianity at low $l$}\label{ltx}
In the derivation of (\ref{clyf}) we made use
of statistics for unordered textures. However, a similar
expression could have been obtained using ordered
textures. In this case one should be wary for the $W^{sl}_n$
are not independent variables. Nevertheless formula
(\ref{clsum}) remains valid if one uses the marginal
distribution functions of $y_n$ in computing $<W^{sl2}_n>$
for each term in (\ref{clsum}). The marginal distribution
functions ${\overline P}_n(y_n)$ are defined as
\begin{equation}
{\overline P}_n(y_n)=\int dy_1... dy_{n-1}dy_{n+1} ... dy_{N_t}
P_{1...N_t}(y_1,..., y_{N_t})
\end{equation}
and using (\ref{pjoint}) they can be found to be
\begin{equation}
{\overline P}_n(y_n)=N(y_n) e^{-M(y_n)}{ M^{n-1}(y_n)\over (n-1)!}
\end{equation}
where $M(y)=\int^y_0 N(x)dx$.
Then:
\begin{equation}\label{clyn}
C^l={<a^2>\over 4\pi}{\sum^{N_t}_{n=1}}
{\int^{y_{ls}}_0} dy_n\, N(y_n)
W^{ls\, 2}(y_n) e^{-M(y_n)}{ M^{n-1}(y_n)\over (n-1)!}
\end{equation}
and since $N_t$ is very large we recover (\ref{clyf}).
The advantage of (\ref{clyn}) is that it allows us to write
\begin{equation}\label{cln}
C^l={\sum_n} C^l_n
\end{equation}
with
\begin{equation}
C^l_n= {<a^2>\over 4\pi}{\int^{y_{ls}}_0} dy_n\, N(y_n)
W^{ls\, 2}(y_n) e^{-M(y_n)}{ M^{n-1}(y_n)\over (n-1)!}
\end{equation}
reporting the contribution of the $n^{th}$
texture to the multipole $C^l$.
\begin{figure}[t]
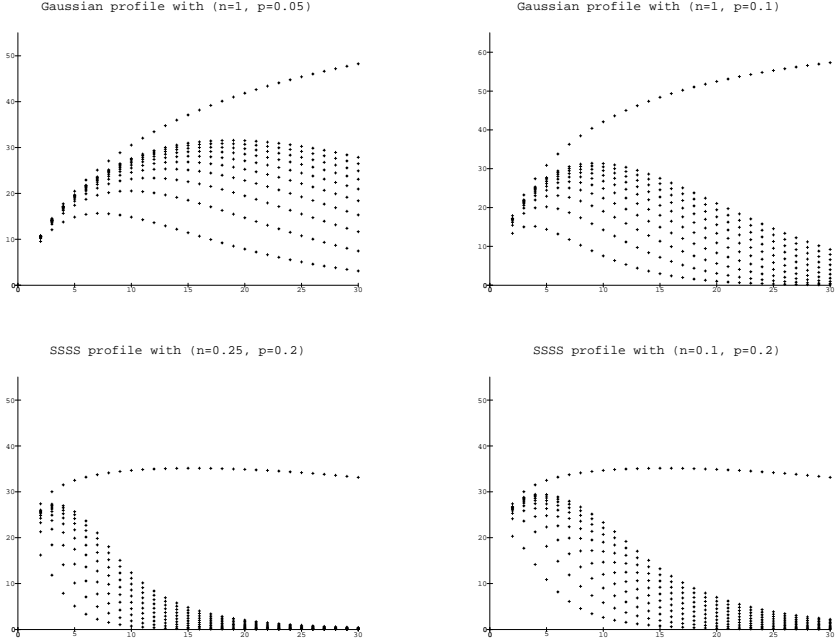

  \begin{center}
    \leavevmode
	{\hbox %
   { \epsfxsize = 6cm  \epsffile{lt005.ps}}
{ \epsfxsize = 6cm  \epsffile{lt01.ps}}
}
	{\hbox %
   { \epsfxsize = 6cm  \epsffile{lt025ss.ps} }
{ \epsfxsize = 6cm  \epsffile{lt01ss.ps}}
}
  \end{center}
\caption{Contribution of the last ten textures for a Gaussian
profile, with $p_s=0.05$ and $p_s=0.1$ ($n=1$ in both cases),
and for a SSSS profile with $n=0.25$ and $n=0.1$ (and $p_s=0.2$).}
\label{ltxeg}
\end{figure}
In Figure \ref{ltxeg} we plotted the $C^{l\star}$ spectrum for four
plausible scenarios: for a Gaussian profile with $p_s=0.05$ and
$p_s=0.1$ ($n=1$ in both cases),
and for a SSSS profile with $n=0.25$ and $n=0.1$ ($p_s=0.2$ for both).
Underneath the spectra we plotted the result of stopping the sum
($\ref{cln}$) at $N_t$ from 1 to 10, thus obtaining the
contribution of the first 10 textures to each $C^l$. In all the cases
considered we notice that the low $l$ multipoles are dominated by
the contribution from the last texture. The $l$ at which this stops
being the case depends on the spot details (profile, $n$, $p_s$),
but $l=2,3$ always seem to be subject to this effect.
Also the intensity of the last texture dominance at low
$l$ depends on the spot details, the effect being notably pronounced
for Gaussian profiles with low $p_s$. As we go up in $l$, the last
texture becomes less prominent, but no other single texture
replaces it as a dominant feature. We may ask how many textures
are responsible for, say, $95\%$
of a given $C^l$ for high $l$. Using the toy hat model
and performing an approximate calculation similar
to the one in Sec.\ref{hrt}, we can show that this number is
proportional to $l^2$. This implies that for high $l$
the number of defects responsible for a given $a^l_m$
is proportional to $l$.

These results have far reaching consequences. If one believes in
flippant applications of the central limit theorem, then
one can expect the $a^l_m$ for high $l<30$ to be independent
Gaussian distributed random variables, as they add up a large
number of independent contributions (proportional to $l$).
We have now shown that this argument can certainly not
be repeated for the low $l$. In fact, our remark in Sec.\ref{apsscl}
that $\sigma^2_{TX}(C_l)>\sigma^2_{Gauss}(C_l)$, and the numerical
results plotted in Fig. \ref{cvcorr},  $prove$  that the
$a^l_m$ are distinctly non-Gaussian at low $l$. This also implies
that these $a^l_m$ are dependent random variables \cite{conf}. We can now
understand better why this is so. The non-Gaussian features
in the $a^l_m$ for low $l$ result from the fact that they are mostly due
to one single lumpy object blatantly different from Gaussian
noise. We may even expect the $a^l_m$ for low $l$ to reproduce
the morphology, size,  and other features of the last defect.
Devising a data-analysis method capable of uncovering these features
is the purpose of \cite{ltx2}. We should stress that both the central limit
theorem and Fig. \ref{cvcorr} strongly suggest, but still
do not prove, that the
$a^l_m$ are Gaussian for high $l<30$.

\begin{figure}[t]
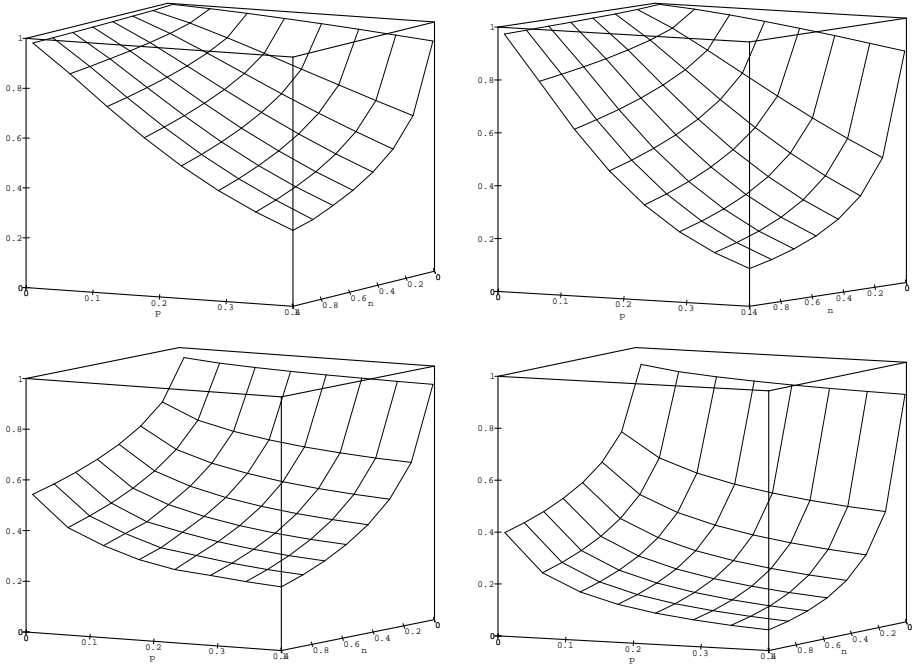

  \begin{center}
    \leavevmode
	{\hbox %
   { \epsfxsize = 6cm \epsffile{quadltx.ps}}
  { \epsfxsize = 6cm \epsffile{octltx.ps}}}
	{\hbox %
   { \epsfxsize = 6cm \epsffile{quadltxss.ps}}
  { \epsfxsize = 6cm \epsffile{octltxss.ps}}}
  \end{center}
\caption{The $\alpha^2$ and $\alpha^3$ function for a Gaussian
(top), and a SSSS profile (bottom).}
\label{clltx}
\end{figure}

The non-Gaussian
toy model studied in \cite{conf} showed in which way the low $l$
multipoles would display non-Gaussian behaviour if they were
fully due to the last texture. The contribution from $n>1$
textures will soften this non-Gaussianity. Hence a plausible
measure for the expected non-Gaussianity present in a low
$l$ multipole is $\alpha^l=C^l_1/C^l$, a quantity which depends
on the profile, $n$, and $p_s$. We have computed numerically
$\alpha^2$ and $\alpha^3$ for a Gaussian and a SSSS profile.
The results are plotted in Fig. \ref{clltx}. A better connection
between $\alpha^l$ and detailed features of non-Gaussianity
measure distribution functions will be given in  \cite{ltx2}.

\section{What's new?}
The focus in this paper was to derive for texture scenarios
whatever one normally derives for Gaussian theories ($C^l$, $C(\theta)$,
and their cosmic variances). We met anomalous behaviour
(eg. formulae (\ref{cov}) and (\ref{sigcorr})) which (theoretically)
indicates non-Gaussianity,
but we have left to the sequels \cite{ltx2}
and \cite{ngsp}  the task
of developing a non-Gaussian data analysis method.
On the whole, within the standard lore, the
results obtained are surprisingly similar to inflationary scenarios.
Still, we found a few novelties.
We discovered that texture spectra are tilted, but not uniformly.
Fitting a tilted spectrum for $l\in(25,30)$ (see Fig. \ref{index})
leaves a significant suppression of power at low $l\in(2,5)$
(see Fig. \ref{clbr1}). We have also
computed directly the cosmic variance
error bars in $C_l$ and $C_{obs}(\theta)$, and found them to be
larger than in comparable Gaussian theories (see formulae
(\ref{cov}) and (\ref{sigcorr}) and Figs. \ref{cvcorr}
and \ref{epsc1}).
In fact, if the non-Gaussian correction is very large
(which does happen for most of the parameter values
suggested in the literature), cosmic variance drowns
the otherwise 1-$\sigma$ suppression of power at low $l$ mentioned
above (see Fig. \ref{clbr1}).
However, one should not give way to despair. Large
non-Gaussian corrections to cosmic variance only hint
that we have not applied to the theory the right data
analysis procedure. To give a flavour of \cite{ngsp}
let us point out that as the $C_l$ are dependent random variables
in non-Gaussian theories (\cite{conf}), it may happen that
{\em the $C_l$ spectrum shape seen by any observer
is never the average $C^l$ spectrum shape}. A more intelligent
method to make predictions on $C_l$ spectra and to estimate
global parameters in the $C^l$ spectrum is in order.
It appears that even a proper study of the $C^l$ (or $C_l$)
spectrum for textures requires non-Gaussian data analysis.

The central result in this paper is undoubtedly the {\em proof}
that texture low $l$ multipoles are strongly non-Gaussian,
and that the last texture is to be blamed for this (see Sec.\ref{ltx}).
We showed how the low $C^l$ are mostly due to the last texture
(slightly perturbed by the one after - see Fig. \ref{ltxeg}),
and so we may expect the last texture's non-Gaussian features
to be imparted on the low $l$ multipoles.
In \cite{conf} we proposed the use of $m$-structure
and inter-$l$ measures to complement the $C^l$.
$m$-structure measures act as multipole shape factors.
Inter-$l$ measures correlate preferred directions
in two multipoles. These two types of quantities
are uniformly distributed
in Gaussian theories. In topological defect scenarios
we can expect low $l$ shape factors to reflect
the morphology of the last defect. Hence their distributions
should be non uniform, peaking at
different values for, say, texture and cosmic
strings. Inter-$l$ measures should also
display the correlations between the various low $l$
which are all due to the same last defect. In remains to
be seen (\cite{ltx2}) how much cosmic confusion there
is between these signals and their Gaussian counterparts.

On a lower key this paper was simply intended as an analytical
model for texture CMBR skies with the greatest possible generality.
We tried to leave as free parameters
whatever simulations have failed to
determine. The point is to find out what exactly
must be decided in order to answer a particular question.
We have concluded, for instance, that the $C^l$ spectrum shape
is insensitive to $n$ or $a$, and depends very little on the spots'
profile, once the identifications  (\ref{ovp}) and (\ref{ova})
have been made. The controlling parameter is ${\overline p}_s$
which fixes the spectrum tilt at intermediate $l$ (see
Fig. \ref{index}). The spectrum normalization, on the other hand,
depends on all the parameters of the model (see Sec.\ref{apscl}),
which must be evaluated
before any prediction for the texture symmetry breaking energy
or bias parameter can be made.  Interestingly enough,
we found that the low $l$ non-Gaussianity effects depend only
on $n$, ${\overline p}_s$, and profile (Sec.\ref{ltx}).
These are not free parameters of the theory, so one ought to decide
on this important issue purely by means of hard work.

The paradigm proposed here cannot easily be adapted to
cosmic string scenarios. Neglecting
non-circular spot modes is,
in texture scenarios, only slightly inaccurate.
The string counterpart
to circular spots, line discontinuities cut-off at horizon
distance (as in \cite{periv}),  is definitely a much
cruder approximation. More importantly,
neglecting texture spot-spot correlations introduces only
a small error (in fact it would not be difficult to
incorporate small correlations in our model). On the contrary,
the various segments of the string network form a Brownian
random walk. Segments linked to each other are therefore
very highly correlated, making an analytical treatment
cumbersome. Uncorrelated string segments (as in \cite{periv})
are not very realistic. Overall, although we can boast
a more realistic set of assumptions
than \cite{periv},  this is mostly
due to the fact that textures are far simpler than strings.
This also justifies why we can go further than \cite{periv}.
We treated with greater rigour  (no ``ergodic'' assumptions)
a much broader set of models. We also managed to  treat
exactly (within the model) the problem of cosmic variance.
Finally,  greater sophistication in the statistical machinery
opened doors to the issue of non-Gaussian data analysis
in defect scenarios (developed further in \cite{ltx2,ngsp}).

\section*{Acknowledgments}
I would like to thank N.Turok and P.Ferreira for discussion
in connection with this paper, and K.Baskerville for
help in its preparation. This work was supported by a fellowship
at St.John's College, Cambridge.

\end{document}